\documentclass[pdflatex,sn-nature]{sn-jnl}

\usepackage{graphicx}%
\usepackage{multirow}%
\usepackage{amsmath,amssymb,amsfonts}%
\usepackage[title]{appendix}%
\usepackage{xcolor}%
\usepackage{manyfoot}%
\usepackage{booktabs}%
\usepackage{siunitx}
\DeclareSIUnit{\wtpercent}{wt.\%}

\usepackage{xifthen}
\newif\iffastfigs
\fastfigstrue   
\theoremstyle{thmstyleone}%
\theoremstyle{thmstyletwo}%
\theoremstyle{thmstylethree}%

\begin{document}

\title[Nanoscale phase mapping in medium-Mn TRIP steel]{Nanoscale mapping of phase-transformation pathways in medium-Mn TRIP steel by multimodal STEM}

\author*[1]{\fnm{Marc} \sur{Raventós-Tato}}\email{marcraven@gmail.com}
\author[1]{\fnm{S. Leila} \sur{Panahi}}\email{spanahi@cells.es}
\author[1]{\fnm{Núria} \sur{Bagués}}\email{nbagues@cells.es}
\author[2]{\fnm{David} \sur{Frómeta}}\email{david.frometa@eurecat.org}
\author[1]{\fnm{Oleg} \sur{Usoltsev}}\email{ousoltsev@cells.es}
\author[2]{\fnm{Núria} \sur{Cuadrado}}\email{nuria.cuadrado@eurecat.org}
\author[1]{\fnm{Joaquín} \sur{Otón}}\email{joton@cells.es}

\affil*[1]{\orgdiv{Methodology Group}, \orgname{ALBA Synchrotron}, \orgaddress{\street{Carrer de la Llum 2–26}, \city{Cerdanyola del Vallès}, \postcode{08290}, \state{Barcelona}, \country{Spain}}}

\affil[2]{\orgdiv{Unit of Metallic and Ceramic Materials}, \orgname{Eurecat, Centre Tecnològic de Catalunya}, \orgaddress{\city{Manresa}, \postcode{08242}, \country{Spain}}}


\abstract{The mechanical response of third-generation advanced high-strength steels is governed by phase transformations at the nanoscale, yet the coupled evolution of chemistry and crystallography remains poorly resolved. Here we apply a correlative STEM approach that enables simultaneous mapping of lattice structure, crystallographic orientation, and phase distribution at \SI{10}{\nano\metre} resolution in a medium-Mn (MMn) TRIP steel. We combine nano-beam electron diffraction (NBED) and energy-dispersive X-ray spectroscopy (EDS) maps to characterize an industrial MMn (\SI{7.15}{\wtpercent} Mn) TRIP steel. Tensile testing of a rolled steel sample was performed, and lamellae were extracted from the resulting deformed and undeformed regions. Mn-resolved EDS provides a chemical fingerprint that, when combined with NBED-based phase segmentation, enables robust ferrite–martensite separation and phase-resolved lattice-parameter refinement. The phase fractions of ferrite, austenite, and martensite are quantified, as well as their corresponding lattices, accompanied by a measurable shift in grain-size distributions and crystallographic texture for the deformed regions. Kernel average misorientation (KAM) maps reveal systematically lower local misorientation in ferrite than in martensite. This multimodal workflow provides a transferable framework for quantitative, phase-resolved analysis of complex multiphase alloys at the nanoscale.}


\maketitle

\section{Introduction}\label{sec:Introduction}
Advanced high-strength steels (AHSSs) are valued in the automotive industry, since they allow lighter vehicles to meet safety standards while improving fuel economy and reducing emissions \citep{kuziak2008ahss,Meknassi2021,Hall2011}. First-generation AHSSs, such as dual-phase (DP) and complex-phase (CP) steels exhibit improved performance by balancing strength and ductility using simple alloying and conventional processing \citep{lesch2017ahss,Knyazeva2013,Fonstein2015,Karelova2009}.

Second-generation AHSSs, particularly twinning-induced plasticity (TWIP) steels, emerged with a focus on maximizing ductility and strain hardening through deformation mechanisms based on mechanical twinning \citep{yu2025review,DeCooman2018,Idrissi2010}. These steels typically contain \SIrange{15}{30}{\wtpercent} manganese (Mn), which stabilizes the austenitic phase at room temperature and facilitates twinning during plastic deformation. As a result, TWIP steels offer exceptional ductility and energy absorption, making them attractive for safety-critical components \citep{guo2024strain,Neu2013}. However, their widespread use has been limited by their low yield strength, high alloying costs, and complex processing requirements \citep{Neu2013,feng2023twip}.

Third-generation AHSSs aim to overcome the limitations of earlier generations by offering a balance between strength, ductility, formability, and cost \citep{Bleck2019BHM3rdGenAHSS,Butuc2024Formability3rdGenAHSS}. This class includes steels based on the transformation-induced plasticity (TRIP) effect, especially medium-Mn (MMn) TRIP steels and quenching \& partitioning (Q\&P) steels \citep{demoor2010strategies,Grajcar2012}. These alloys rely on austenite stabilization and its controlled transformation into martensite during deformation to enable high strain hardening and energy absorption \citep{sun2019yielding,steineder2017microstructural}. The addition of Mn not only stabilizes austenite during quenching and heat treatment but also helps retain sufficient amounts of this metastable phase for effective TRIP behavior during forming operations \citep{benzing2019mechanical,zheng2018formability,Shen2015}.

Despite their promising mechanical performance and favorable strength-to-cost ratios, MMn TRIP steels still face significant challenges related to formability and welding, especially for complex-shaped automotive components \citep{mohapatra2025advances,Billur2019}. These issues are closely linked to the underlying microstructural evolution during deformation and processing, making high-resolution crystallographic characterization invaluable for capturing phase-dependent deformation mechanisms \citep{decooman2013tem,hung2021grain,brodusch2021semtem,Petrov2007}. To address this microstructural characterization in MMn TRIP steels, we focused on a correlative approach that combines different advanced electron microscopy techniques.

\begin{figure*}[htbp]
\centering
\includegraphics[width=1\textwidth]{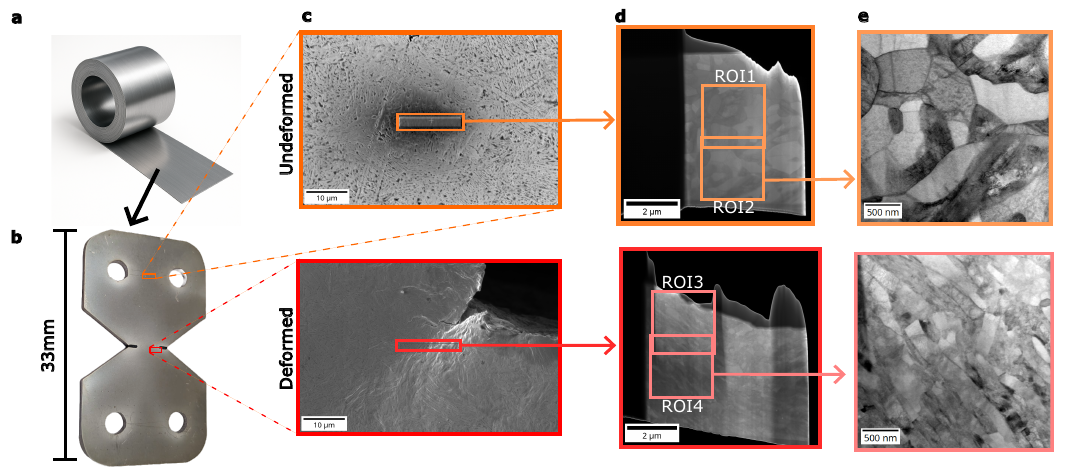}
\caption{\textbf{Workflow from sample preparation to selection of ROIs}.
\textbf{a} Schematic of the MMn1000 industrially rolled medium-manganese steel sheet. 
\textbf{b} Double-edge notched tensile (DENT) specimen extracted transverse to the rolling direction. 
\textbf{c} Scanning electron microscope images of the undeformed and deformed regions selected for TEM lamella preparation. 
\textbf{d} Dark-field scanning transmission electron microscopy (STEM) images of the lamellas, showing ROIs identified for further correlative analysis. 
\textbf{e} Corresponding bright-field STEM images indicating two of the ROIs studied in this work, corresponding to ROIs 2 and 4.}\label{fig:Dogbone}
\end{figure*}

A double-edge notched tensile (DENT) specimen was extracted from an industrially rolled MMn1000 steel grade sheet and tested under tension, as illustrated in Figure~\ref{fig:Dogbone}. From this sample, two lamellae for the transmission electron microscope (TEM) were prepared using the focused ion beam-scanning electron microscope (FIB-SEM): one lamella from an undeformed area and another one from a heavily deformed area. Four regions of interest (ROI), two deformed and two undeformed, were selected for subsequent analysis. These ROIs were examined using a combined methodology of scanning transmission electron microscopy (STEM) techniques: nano-beam electron diffraction (NBED)\citep{beche2013strain,ophus2019four} and energy-dispersive X-ray spectroscopy (EDS) \citep{hodoroaba2020energy,Allen2011STEMEDS,Lugg2015QuantitativeEDS}. While STEM-EDS provided local compositional contrast based on Mn atomic concentration, STEM-NBED enabled high-resolution crystallographic phase and orientation mapping. This correlative strategy allowed us to reconstruct microstructural changes down to \SI{10}{\nano\metre} resolution and to resolve ferrite from martensite in terms of their $c/a$ lattice parameter ratio.

\section{Results}\label{sec:Results}

\begin{figure*}[htbp]
\centering
\includegraphics[width=1\textwidth]{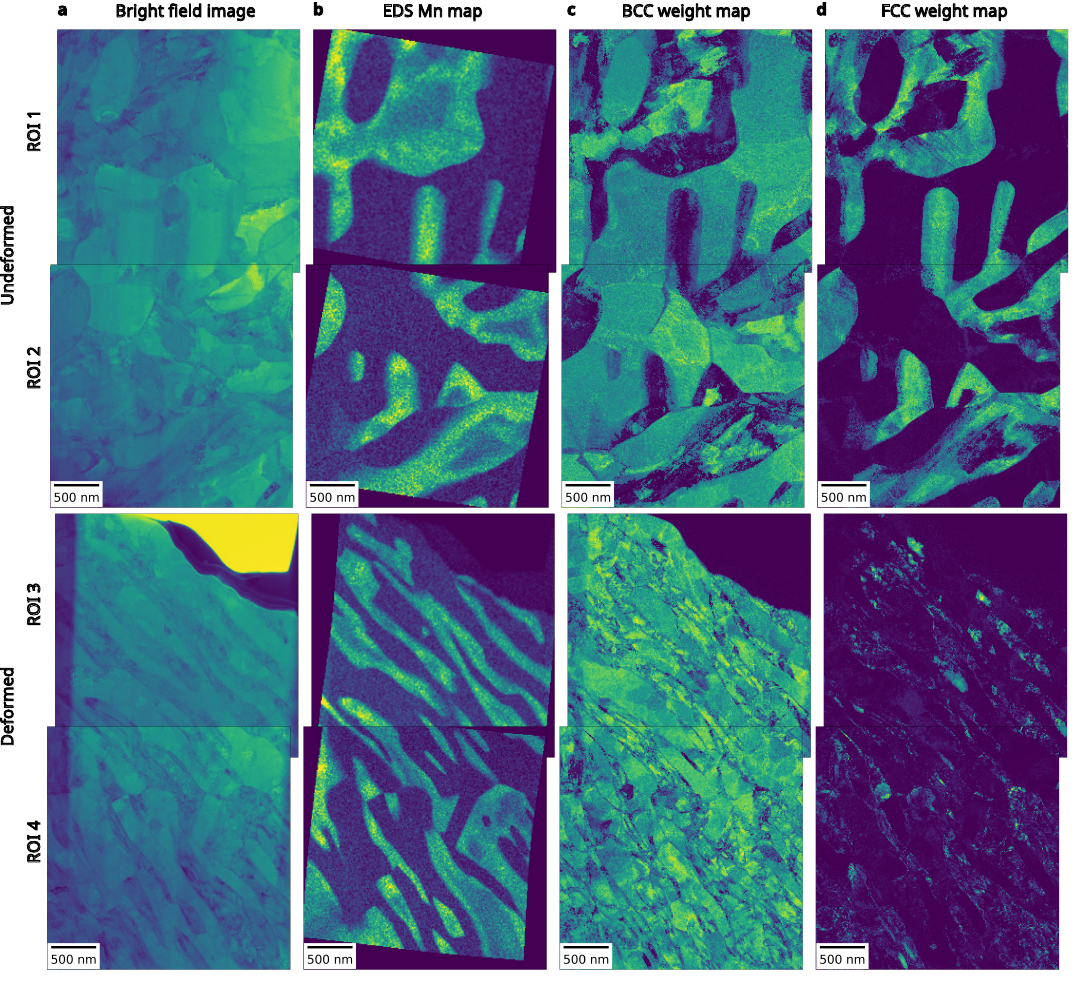}
\caption{\textbf{Bright field images, spectral maps and phase weights}. \textbf{a} Bright-field images showing intensity variations in electron transmission. 
\textbf{b} Energy-dispersive X-ray spectroscopy (EDS) signals corresponding to the Mn peaks aligned with the nano-beam electron diffraction data through affine transformations. 
\textbf{c} and \textbf{d} Map of the relative concentrations of body-centered cubic (BCC) and face-centered cubic (FCC) phases. There is a spatial overlap between ROIs 1 and 2 (undeformed lamella) and ROIs 3 and 4 (deformed lamella).}
\label{fig:EDS}
\end{figure*}

After selecting the Mn energy peaks of the EDS and determining, for each pixel, the relative crystal-structure fraction (BCC-ferrite ($\alpha$) vs FCC-austenite ($\gamma$)), the results can be visualized for the four ROIs in Fig.~\ref{fig:EDS}. For each ROI, Fig.~\ref{fig:EDS} shows a bright-field (BF) image, an EDS map of the Mn K$\alpha$ signal, and complementary maps of the relative concentrations of BCC and FCC obtained by processing the NBED data. A slight misalignment arises due to the use of different optical configurations for EDS and NBED; its correction is detailed in the \nameref{sec:Methods} section.

\color{black}

A first observation is that phase boundaries in all ROIs follow the grain boundaries seen in the BF images (as expected). The EDS and phase maps provide context for the transformations taking place in both regions. Mn acts as an austenite stabilizer in steel, so the EDS and FCC maps of the undeformed region match almost perfectly. In the deformed sample, the Mn-rich regions show a clear preferential orientation (caused by plastic deformation) that aligns with features in the BF image. In this case, there is little correlation between the EDS and the BCC/FCC maps, as the austenite has almost completely transformed into martensite, which appears together with ferrite in the BCC map.Because deformation-induced martensitic transformation is displacive, no long-range redistribution of substitutional Mn is expected on the relevant timescale. Recent experimental studies on medium-Mn steels confirm that Mn heterogeneities persist during room-temperature deformation and that martensite largely inherits the Mn content of the parent austenite \citep{Christian_TheoryTransformationsMetalsAlloys_2002,Yang_Wu_Yi_2019_ReverseTransformation_MediumMn,sun2021natmat,PorterEasterlingSherif_PhaseTransformations_2022,zhang2023core}.

\begin{figure*}[htbp]
\centering
\includegraphics[width=1\textwidth]{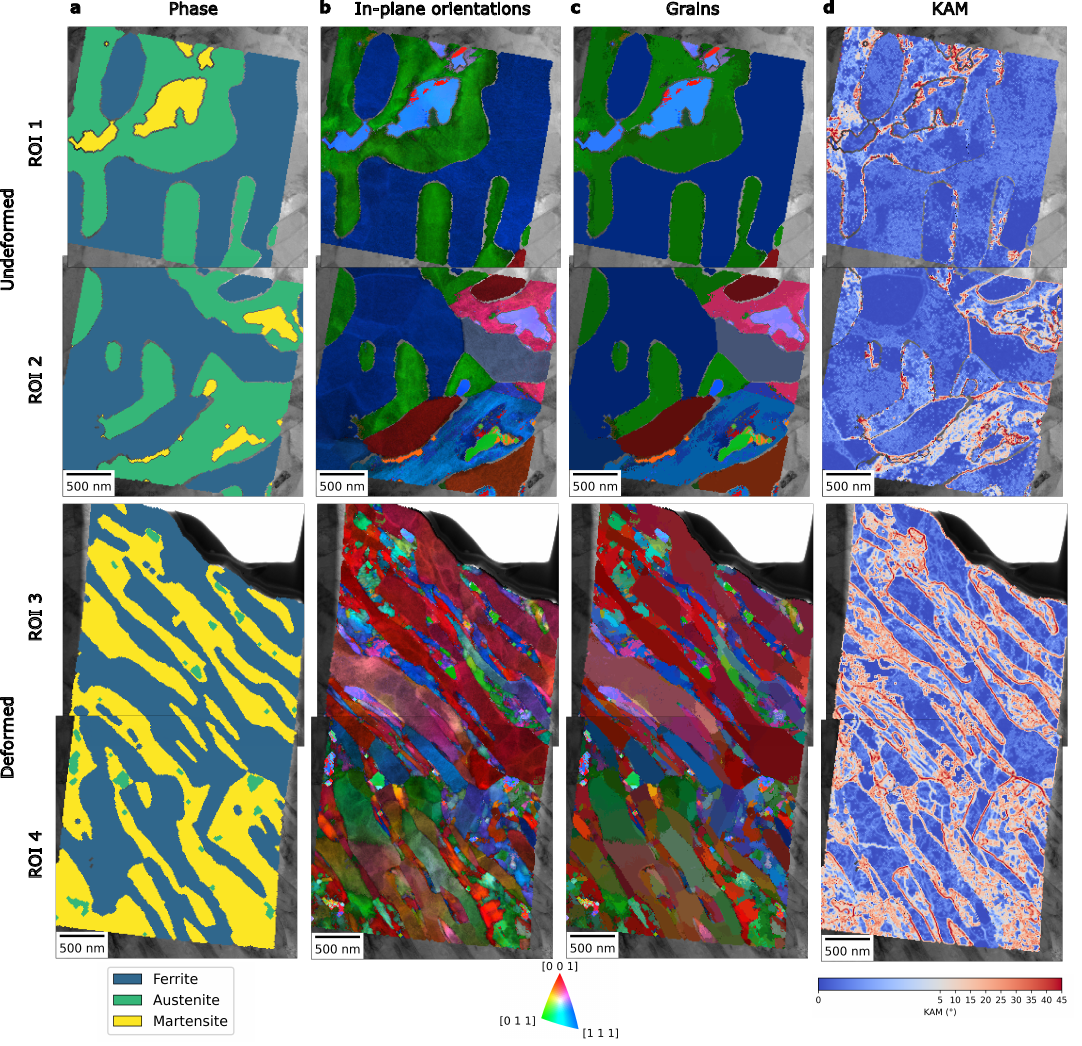}
\caption{\textbf{Phase, orientation, grain and misorientation maps}.
\textbf{a} Phase maps showing the spatial distribution of ferrite, austenite, and martensite, generated by overlap of EDS and phase maps from Fig.~\ref{fig:EDS}.
\textbf{b} In-plane crystal orientations maps with inverse pole figure indicating the lattice plane with respect to the vertical direction. 
\textbf{c} Grain maps obtained by clustering neighboring pixels from the orientation map with misorientations below \SI{5}{\degree}, illustrating the grain morphology for each phase. 
\textbf{d} Kernel average misorientation (KAM) maps showing the average misorientation angle of each pixel with respect to its eight closest neighbors. The KAM colormap spans \SIrange{0}{45}{\degree} and is centered at \SI{5}{\degree} (visualization only) to enhance discrimination between low-angle and high-angle boundaries.}\label{fig:Merit}
\end{figure*}

Figure~\ref{fig:Merit}\textbf{a} shows the phase map built from the maps in Figure~\ref{fig:EDS}\textbf{b,c,d} (see the procedure in the \nameref{sec:Methods} section). The total phase fractions in the undeformed region (ROIs 1 and 2) are \SI{55}{\percent} ferrite, \SI{37}{\percent} austenite, and \SI{8}{\percent} martensite. In the deformed region (ROIs 3 and 4), the fractions are \SI{54}{\percent} ferrite, \SI{2}{\percent} austenite, and \SI{44}{\percent} martensite.

The orientation maps (Fig.~\ref{fig:Merit}b) show similar in-plane orientations for adjacent austenite and ferrite regions, consistent with structurally coherent and closely intergrown domains. The corresponding grain and kernel average-misorientation (KAM) maps are shown in Fig.~\ref{fig:Merit}c,d.

The undeformed ROIs 1 and 2 show micrometer-sized domains of austenite and ferrite in dendritic structures with small islands of martensite in the Mn-rich regions. The deformed ROIs 3 and 4 show a highly textured distribution of ferrite and martensite grains with small remaining islands of austenite in Mn-rich regions. Ferrite exhibits a significant grain size decrease relative to the ferrite in the undeformed ROIs. By contrast, the martensitic region displays significant microstructural complexity: a dense network of fine, needle-like grains is embedded within the Mn-rich matrix, with characteristic dimensions of just a few tens of nanometers. The high degree of overlap between the Mn-rich regions of the EDS and the nanocrystalline martensitic structures in the orientation and grain maps supports that Mn migration is negligible during deformation and thus serves as a fingerprint to segment ferrite from martensite in medium and high Mn steel alloys. KAM maps, which are commonly used as a semi-quantitative measure of local lattice curvature and deformation heterogeneity, show systematically higher values in martensitic regions than in ferrite. This behavior is consistent with transformation-induced lattice distortion in martensite compared to dislocation-mediated plasticity in ferrite\citep{Wilkinson2006KAM,Calcagnotto2010KAM,wilkinson2019gnd,mohapatra2024kam}.

Upon deformation, the microstructure undergoes significant reorganization. All regions become visibly elongated due to the applied tensile strain, and the residual austenitic domains are now small and sparsely distributed. The Mn-depleted ferrite shows increased mosaicity, with a broader spread of orientations and several distinct grains across the region. The Mn-rich areas, now largely transformed into martensite, exhibit a fragmented microstructure composed of nanocrystalline domains with a wide orientation spread\citep{,Morito2006Martensite}. These distinctions between Mn-rich and Mn-depleted regions are now evident even in the absence of explicit phase labelling, based on differences in grain size, shape, and misorientation contrast. On this basis, the phase-segmented regions were subsequently used to independently calibrate the crystal lattices of ferrite, austenite, and martensite, yielding the phase-resolved lattice parameters discussed below.

\begin{figure*}[htbp]
\centering
\includegraphics[width=1\textwidth]{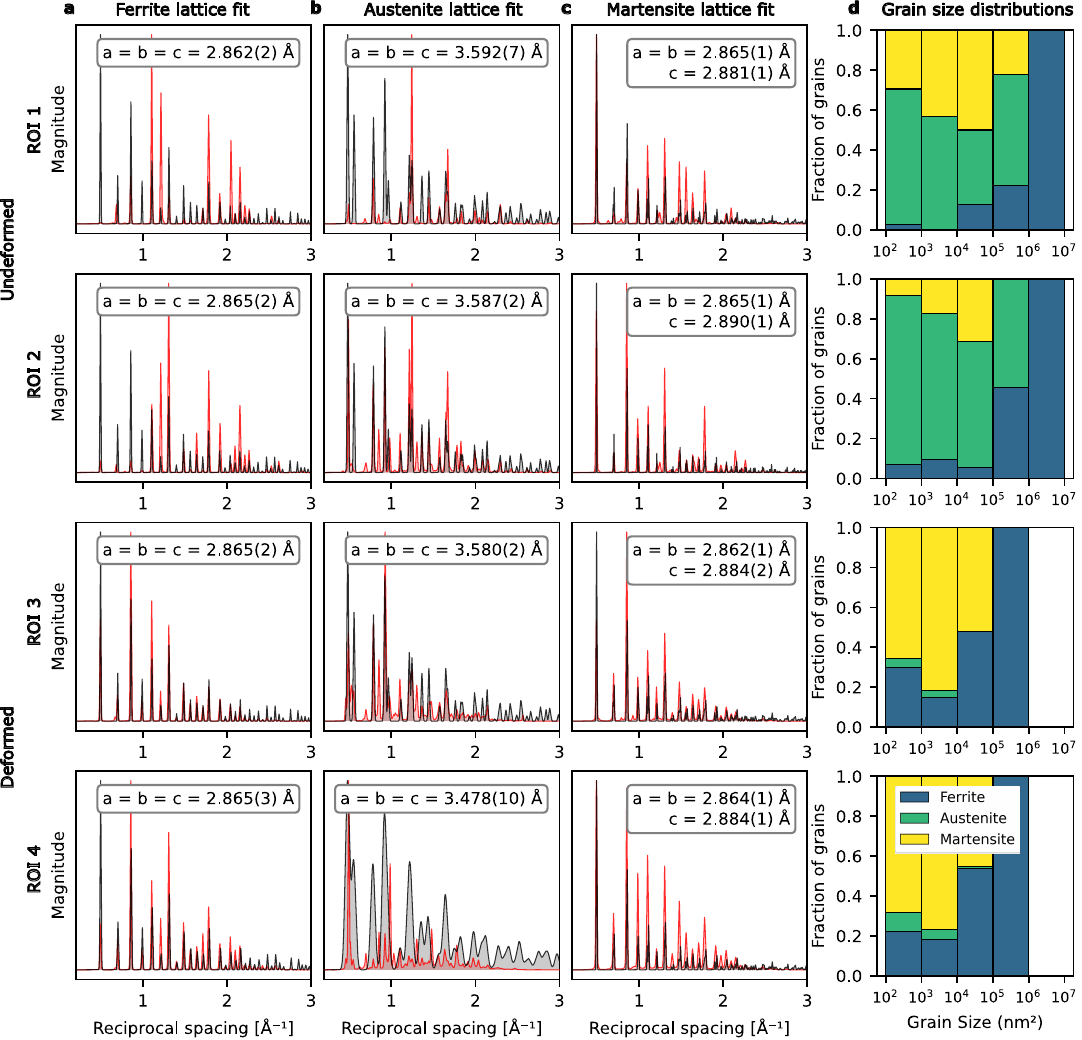}
\caption{\textbf{Quantitative analysis for each phase of the ROIs}.
\textbf{a} Lattice calibration of the ferrite regions (BCC).
\textbf{b} Lattice calibration of the austenite regions (FCC).
\textbf{c} Lattice calibration of the martensite regions (BCT).
\textbf{d} Grain size distributions for each of the phases (legend in ROI 4) obtained from the clustered grain maps from Fig.~\ref{fig:Merit}c .}\label{fig:Rietvelds}
\end{figure*}

The results of the crystal lattice calibrations for each phase are shown in Fig.~\ref{fig:Rietvelds}{a,b,c}. The lattice parameters were obtained from phase-resolved 1D diffraction patterns, using the same calibration procedure as for the Au–Pd reference described in the \nameref{sec:Methods} section. The values in parentheses indicate the standard deviation (estimated from the covariance matrix) for the last significant digit of each lattice parameter. The grain size distributions shown in Fig.~\ref{fig:Rietvelds}{d} were obtained from the grain clustering analysis in Fig.~\ref{fig:Merit}c and represent phase-resolved grain size statistics.

To test whether the diffraction experiment was sensitive enough to distinguish the slightly tetragonal structure of martensite from the cubic structure of ferrite, all regions were first fit using the same parameters (with the \( c \) lattice parameter allowed to vary). The resulting lattice parameters are shown over the 1D diffraction patterns in Figure~\ref{fig:Rietvelds}. The standard deviations in parentheses were about one order of magnitude higher for the ferrite regions when \( c \) was free to vary. This large uncertainty came from numerical instability in the non-linear least-squares fit (caused by strong correlation between two free parameters). This effect was not observed in the martensite regions, which showed uncertainties on the order of \( \SI{2e-3}{\AA} \). When the ferrite regions were fit with a cubic lattice constraint (as it is in Fig. \ref{fig:Rietvelds}\textbf{a}), the instability disappeared and the errors decreased to values similar to those of martensite.

Quantitatively, ferrite and austenite show similar grain sizes in the undeformed condition (with ferrite containing the largest grains). Martensite appears only in small isolated regions within the undeformed ROIs, since the spatial overlap between Mn enrichment and the FCC phase is limited. After deformation, the transformation of austenite into martensite causes a strong reduction in the average grain size of both phases (most noticeably in austenite, whose fraction becomes residual). Ferrite retains some of its large grains (especially those previously intergrown with austenite) and shows a clear preferential orientation, although its overall average grain size still decreases.

\section{Discussion}\label{sec:Discussion}

We present a nanoscale-resolved phase, orientation, and grain mapping of a MMn TRIP steel using correlative STEM measurements of NBED and EDS at \SI{10}{\nano\metre} resolution. MMn TRIP steels such as MMn1000 are ideal test systems due to their nanometric texture and chemical segregation. These features enable direct combination of EDS chemical maps and NBED diffraction data.

In undeformed regions, grain sizes exceed the lamella thickness, minimizing overlap. In deformed areas, finer grains cause significant phase and grain overlap within single probe positions. Template-matching ACOM approaches typically assume a single dominant orientation per probe position. In nanocrystalline or heavily deformed regions, however, the interaction volume may encompass multiple grains, leading to overlapping diffraction patterns and reduced indexing reliability—a known limitation of conventional ACOM methodologies \citep{valery2017overlap,rauch2021acom}.

The robustness of the analysis is confirmed by the close agreement between the independently processed deformed ROIs (3 and 4), which show consistent phase, orientation, grain, and KAM maps. Although demonstrated here for Mn-partitioned TRIP steels, the workflow is broadly applicable to other alloyed steels. Future work should integrate this nanoscale approach with mesoscale 2D synchrotron XRD mapping to connect bulk diffraction metrics with local nanostructure.  

The workflow links local chemistry, lattice distortion, and phase evolution during deformation in TRIP steels. In combination, EDS and NBED offer a reproducible framework for resolving phase-transformation pathways in multiphase alloys with nanometric heterogeneity. By combining diffraction-based analysis with compositional contrast, it enables quantitative assessment of phase stability and transformation kinetics at the nanoscale. Extending this methodology to mesoscale diffraction or \textit{in-situ} loading experiments will allow correlation of local transformations with macroscopic mechanical behavior.

\section{Methods}\label{sec:Methods}

\subsection{Sample preparation and data acquisition}

Two TEM lamellae were prepared from a mechanically tested DENT specimen of a \SI{1.5}{\milli\metre} thick industrially hot-rolled MMn1000 steel manufactured by \textit{Voestalpine AG}. The nominal chemical composition of the alloy was \SI{0.099}{\wtpercent} C, \SI{0.56}{\wtpercent} Si, and \SI{7.15}{\wtpercent} Mn. One lamella was extracted from a region near the tensile grips that experienced negligible deformation, while the other was taken from the plastically strained zone adjacent to the notch tip. These deformed and undeformed regions of the DENT specimen are shown in Figure~\ref{fig:Dogbone} c.

The tensile test was carried out in a \textit{Zeiss Ultra Plus} field-emission scanning electron microscope (FE-SEM) equipped with a \SI{5}{\kilo\newton} Deben microtensile stage. After testing, cross-sectional lamellae were prepared and thinned using a Ga$^{+}$ focused-ion beam (FIB, Helios 5 UX, Thermo Fisher Scientific) operated between \SI{30}{\kilo\volt} and \SI{2}{\kilo\volt}. Low magnification dark-field (DF) images were collected with a STEM detector installed in the FIB-SEM. From each lamella, a $\SI{2.7}{\micro\metre} \times \SI{2.7}{\micro\metre}$ area was selected for high-resolution multimodal STEM analysis combining EDS and NBED.

All STEM measurements were performed at the Joint Electron Microscopy Center at ALBA (JEMCA) using a Thermo Fisher Scientific Spectra 300 double-corrected STEM operated at \SI{300}{\kilo\electronvolt}. STEM-EDS data was collected using a convergence semi-angle of \SI{19.5}{\milli\radian}, with a BF detector and a High-Angle Annular Dark-Field detector for STEM imaging, and a Super-X detector for EDS. STEM-NBED data was collected using a convergence semi-angle of \SI{0.2}{\milli\radian} and an Electron Microscope Pixel Array Detector (EMPAD). BF images from the STEM-NBED datasets were obtained using a virtual circular detector mask including only the zero order. EDS maps were acquired over a \(512\times512\) grid with a step size of \SI{5.27}{\nano\metre}, while NBED maps were acquired at 256×256 positions with a step size of \SI{10.55}{\nano\metre}.

\subsection{Instrument Calibration}

\begin{figure*}[htbp]
\centering
\includegraphics[width=0.5\textwidth]{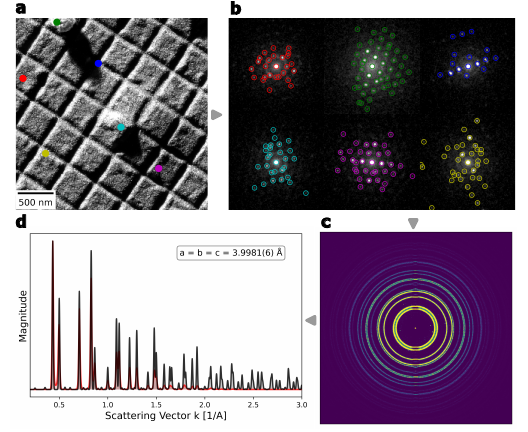}
\caption{
\textbf{Reciprocal space calibration with Au-Pd}.
\textbf{a} Virtual dark-field image of the calibrant with 6 randomly selected pixels highlighted.
\textbf{b} Diffraction patterns from the selected pixels and Bragg peaks harvested.
\textbf{c} Virtual Bragg map combining all the peaks harvested from the calibrant dataset.
\textbf{d} Integrated 1D diffraction pattern from c with the best lattice fit found during calibration.}\label{fig:Calibration}
\end{figure*}

To determine the reciprocal pixel size and diffraction resolution of the system, a Au-Pd calibrant sample was measured right after the steel data acquisition. Using the \texttt{py4DSTEM} Python package \citep{savitzky2021py4dstem}, Bragg peaks were extracted from diffraction patterns, corrected for beam shift and elliptical distortion, and radially integrated. The resulting 1D pattern was then fitted to the known lattice constants of Au-Pd in Figure~\ref{fig:Calibration}.

Fig.~\ref{fig:Calibration} shows a virtual DF image (integrating the STEM-NBED data except the zero order) of the Au-Pd calibrant with Bragg peaks mapped at different regions of the sample using the \texttt{py4DSTEM} Python package. Next to it, a virtual Bragg map upsampled 10 times to 1280x1280 pixels, and the resulting 1D integrated pattern with the result of the reciprocal pixel calibration, obtaining a reciprocal pixel size of \( \SI{0.03397}{\AA^{-1}} \). 

The lattice of the Au-Pd calibrant was refined to establish a baseline for the measurement uncertainty associated with the NBED-1D integration methodology. Assuming cubic symmetry, with all lattice parameters constrained to be equal, the non-linear least-squares refinement yielded a lattice parameter with a standard deviation of \( \SI{6e-4}{\AA} \). This uncertainty is an order of magnitude smaller than those obtained for the actual samples presented in Fig.~\ref{fig:Rietvelds}, as expected for a well-characterized calibrant reference material.

\subsection{Correlative EDS–NBED}\label{sec:DataFusion}

Because the collection mode of EDS is in nanprobe mode while NBED data is in microprobe mode, there is a relative rotation between both datasets. Image rotation and misalignment were corrected by computing affine transformations using fiducial landmarks within the BigWarp ecosystem of ImageJ \citep{Schroeder2021ImageJ}. The EDS maps were rotated into the NBED coordinate frame for simplicity (rotating the STEM-NBED dataset would require more complex interpolation). The rotated EDS maps were then binarized using Otsu thresholding to separate Mn-rich and Mn-depleted regions.  

Phase segmentation into ferrite, martensite, and austenite was based on the integration of diffraction-based crystallographic classification and compositional contrast from EDS. Automated crystal orientation mapping (ACOM) from \textit{py4DSTEM} \citep{ophus2022acom} was applied to the NBED data to determine the most likely crystal orientation at each probe position, using ferrite and austenite reference structures for the initial estimation of the phase weights (Fig.~\ref{fig:EDS}c,d). The ACOM algorithm determines crystal orientations by matching experimental diffraction peak positions (see Fig.~\ref{fig:Calibration}b) with simulated patterns generated from known lattice parameters and atomic positions. A grid of simulated patterns (the so-called orientation plan) was computed with a \SI{1}{\degree} step size for both in-plane and out-of-plane rotations, within the orientation space defined by the crystal symmetry. For all cubic systems, sampling orientations within the region bounded by the [001], [011], and [111] directions is sufficient due to their high symmetry.

Experimental and simulated peak patterns were compared in Fourier space, and the orientation yielding the highest correlation value was selected as the best-fit solution. The likelihood of each pixel belonging to austenite or ferrite was then estimated using the phase quantification routine in \textit{py4DSTEM}, which compares correlation values and peak intensities for each phase.

Final phase identification was performed by combining the crystallographic and compositional masks. Pixels classified as BCC and Mn-rich were assigned to martensite, while those classified as BCC and Mn-depleted were assigned to ferrite. All pixels identified as FCC (independent of Mn content) were assigned to austenite. Empirically, all FCC regions were fully contained within the Mn-rich domain, confirming the consistency of the segmentation and the strong spatial correlation between Mn enrichment and FCC stability.

Following final phase assignment, orientation-resolved and lattice-calibration analyses were carried out independently for ferrite, austenite, and martensite using the corresponding phase masks. Grain maps (Fig.~\ref{fig:Merit}c) were generated in \textit{py4DSTEM} by clustering neighboring pixels with mutual misorientations below $\SI{5}{\degree}$, which defines the grain separation threshold used throughout this work. From this, grain size statistics were computed separately for each crystallographic phase and shown in Fig.~\ref{fig:Rietvelds}{d}.

Kernel average misorientation (KAM) maps (Fig.~\ref{fig:Merit}d) were computed using a custom Python implementation adapted from the \textit{MTEX} formalism \citep{bachmann2010texture}. For each pixel, the average misorientation with respect to its eight nearest neighbors was calculated within the same phase mask, thereby suppressing artificial misorientation contributions arising from phase interfaces. 

For visualization purposes only, the KAM colormap was centered at \SI{5}{\degree} (without modifying the underlying KAM values), enhancing contrast between low-angle misorientations associated with ferrite and higher-angle misorientations typical of martensitic regions.

\subsection{Large language model use}

The GPT-5 model has been used for syntax proofing and organization of the paper, to generate Fig.~\ref{fig:Dogbone}a and for debugging the python code used for analyzing the data and generating the figures. The model was not used to generate scientific interpretations or conclusions.

\section{Acknowledgements}\label{sec:Acknowledgements}

The authors gratefully acknowledge all members of the Sup3rForm consortium for their contributions and collaboration. The consortium includes: Fundació Eurecat (Spain), ALBA-CELLS (Spain), Voestalpine Stahl GmbH (Austria), ArcelorMittal Maizières Research SA (France), Centro Ricerche Fiat SCPA (Italy), MA SRL (Italy), Luleå Tekniska Universitet (Sweden), and Aalto Korkeakoulusäätiö SR (Finland).

\section*{Declarations}

\begin{itemize}
\item Funding

This work has been funded by the Sup3rForm project, which has received support from the European Union’s Research Fund for Coal and Steel (RFCS) under project number 101112540. M.R., S.L.P., N.B., and O.O. acknowledge additional support from the Advanced Materials programme, funded by the Spanish Ministry of Science and Innovation (MCIN) through the European Union’s NextGenerationEU (PRTR-C17.I1), and by the Generalitat de Catalunya.

\item Conflict of interest

The authors declare no competing interests.

\item Ethics approval and consent to participate

Not applicable

\item Consent for publication

Not applicable

\item Data availability 

The datasets generated and analyzed during this study will be made available on Zenodo.

\item Materials availability

Not applicable

\item Code availability 

 The corresponding analysis workflows will be published as Jupyter notebooks in an online repository.
 
\item Author contribution

\noindent
D.F. coordinated the project. O.U. and J.O. designed the experimental procedure and organized the beamtime. N.C. prepared the samples and performed the tensile tests. S.L.P. and N.B. extracted the lamellae and performed the TEM measurements. M.R. performed the data analysis and wrote the paper. All authors provided discussion and comments.
\end{itemize}

\bibliography{sn-bibliography}

\end{document}